# NEEDMINING: IDENTIFYING MICRO BLOG DATA CONTAINING CUSTOMER NEEDS

*Research*

Kuehl, Niklas, Karlsruhe Institute of Technology, Germany, kuehl@kit.edu

Scheurenbrand, Jan, Karlsruhe Institute of Technology, Germany, jan.scheurenbrand@kit.edu

Satzger, Gerhard, Karlsruhe Institute of Technology, Germany, gerhard.satzger@kit.edu

## Abstract

*The design of new products and services starts with the identification of needs of potential customers or users. Many existing methods like observations, surveys, and experiments draw upon specific efforts to elicit unsatisfied needs from individuals. At the same time, a huge amount of user-generated content in micro blogs is freely accessible at no cost. While this information is already analyzed to monitor sentiments towards existing offerings, it has not yet been tapped for the elicitation of needs. In this paper, we lay an important foundation for this endeavor: we propose a Machine Learning approach to identify those posts that do express needs. Our evaluation of tweets in the e-mobility domain demonstrates that the small share of relevant tweets can be identified with remarkable precision or recall results. Applied to huge data sets, the developed method should enable scalable need elicitation support for innovation managers—across thousands of users, and thus augment the service design tool set available to him.*

# 1     Introduction

Identifying the needs of (potential) customers is an important task in order to improve or design new market-driven services (Goldstein et al., 2002; Herrmann et al., 2000; Teare, 1998). Typically used methods in the fields of Marketing or Service Design include observations, surveys and interviews (Edvardsson et al., 2012; Hauser and Griffin, 1993), applied both in academic research (Driscoll, 2011) and in industry (Kärkkäinen and Elfvengren, 2002). However, those methods are time-consuming (Hauser and Griffin, 1993) and cost-intensive (Fisher et al., 2014), as they do not scale across larger numbers of potential customers. In order to provide an artifact enabling managers to identify the needs of (potential) customers based on micro blog data, we are conducting a larger Design Science Research (DSR) project following the guidelines of Kuechler and Vaishnavi (2008) . In this "Needmining" project, we explore an alternative approach of eliciting and gathering customer needs: The smart use of free and publicly available micro blog data (e.g. Twitter) could replace or augment existing approaches. In micro blogs, individuals express their needs, for example, in a moment of dissatisfaction or inconvenience, which points to an underlying demand for a certain (service) solution. As a first step of our overall Needmining project, we build an artifact to classify micro blog data to separate out those entries that contain needs. In this paper, we present the work of this first step of our overall DSR project. With an accurate and well-trained classification artifact, we will be able to identify need information from very large, unseen data sets in the future. This will enable us to screen and elicit needs of thousands of (potential) customers—without conducting time-consuming and cost-intensive surveys or interviews. Subsequently, the identified posts can be further analyzed to extract and specify concrete needs.

The work described in this paper represents an exaptation (Gregor and Hevner, 2013) as it extends Machine Learning approaches to the new problem of scalable needs elicitation in micro blog data. Thereby, our paper contributes to the body of knowledge as we (1) propose the idea of Needmining from micro blog data, (2) provide a concrete analytical method to identify relevant micro blog posts, and (3) prove within a concrete domain (e-mobility) that "automated" Needmining is well feasible via our classification method. Applied to huge data sets, the developed classification tool should already enable scalable need elicitation support for innovation managers. Needmining can be a step change for the development of new services as it allows to both automatically and continuously screen huge groups for needs (efficiency) and to not miss out on latent needs (effectiveness) in the future. With our work, the identified "need tweets" will still have to be analyzed by humans. Future work will then be devoted to further ease that task by eliciting and classifying more comprehensive need information—beyond the mere identification of relevant tweets.

The remaining paper is structured as follows: Section 2 positions the paper within related work and clarifies key terminology. Our method and study implementation is outlined in section 3, before our results are summarized in section 4. Section 5 concludes with a short summary, a discussion of limitations of the work as well as managerial and theoretical implications.

# 2     Terminology and Related Work

Before we suggest and apply a method to screen micro blog data for need information, we first want to specify the term "need", explain micro blog data, and choose a domain to test our approach.

## 2.1     Customer Needs

To understand the customer's requirements and ways to satisfy them is one of the most important challenges in marketing. Ulwick and Bettencourt (2007) show which characteristics of requirements are essential to turn them into new services, while Baida et al. (2005) develop an ontology of customer needs and their connection to possible service offerings. Over the last decades, the analysis of "need finding" has been a key ingredient in Design Thinking and Service Design methodologies (Feldmann and Cardoso, 2016). Kotler and Armstrong (2001) separate customer requirements into three distinct categories: *Needs* are defined as "states of felt deprivation", which include the most basic human requirements for humans like food and shelter. *Wants* are people's desires for specific satisfiers of needs that are shaped by society and personal preferences. When wants (for a specific product or service) are backed by the ability to buy them, they are called *demands*: A person might have the need for mobility, wants a car, and demands a Volkswagen Golf. Needs are often intangible—for example, the needs for mobility or financial security can be interpreted and satisfied in many ways. Therefore, individuals concretize them implicitly by transforming them into wants and demands. For the purpose of this work—identifying micro blog data containing customer need information—there is little to be gained differentiating between needs, wants, and demands: In a first step, any information, regardless of the level of granularity, is valuable information for a marketing or innovation manager. For simplicity, we, therefore, stick with the term *customer need*—taking all three mentioned types (needs, wants, and demands) into account.

### 2.2 Micro Blog Data and Their Analysis

Micro blogs allow users to share content in short messages with the public. Micro blogging services are designed in such way that users can conveniently post content "on the go" via smartphones (Gaonkar et al., 2008). Motivations of individuals to intensively use micro blogging services are analyzed in Java et al. (2007), listing various intentions of different communities. The most frequently used micro blog providers are SINA Weibo—being very popular in China (Guo et al., 2011)—as well as Twitter. Twitter is the most popular micro blog service (Deutsch, 2016) as 316 million daily active users[1] around the globe post 500 million user messages per day. We, therefore, select Twitter as the micro blog database for this work.

Technical approaches to distill information from micro blogs draw on the fields of Predictive Analytics, in particular Natural Language Processing (NLP), and Text Analysis: In the field of NLP in combination with Machine Learning techniques, the basic concept of processing a natural language originates from Turing (1950). Turing states that computers need to descramble, process, and understand natural languages in order to act and think like a human. Concerning the basic concepts of analyzing (large sets of) text data in general, a great amount of literature is available: Manning and Schütze (2000) describe the foundations of statistical NLP and propose practical tools. Feldman and Sanger (2007) show advanced approaches of such tools in the area of "Text Mining" —a term which had previously been introduced by Feldman and Dagan (1995). In a recent study from Evans (2014), the researcher develops an analytical approach for the analysis of large textual data sets. Concerning the analysis of micro blog data in particular, key contributions on the retrieval of opinions (sentiment analysis) can be found in Ku et al. (2006) and Luo et al. (2012). A user modeling and personalization approach of such data is presented by Abel et al. (2011). For more specific, topic-centered approaches, Zhao et al. (2011) suggest an effective method for topical key phrase extraction, while Meng et al. (2012) analyze how to summarize opinions for certain entities (like celebrities and events) with an entity-centric and topic-oriented opinion mining approach. Recent studies also show the identification (Alsaedi et al., 2014) and prediction (Boecking et al., 2014) of disruptive events using micro blog data. However, to the best of our knowledge, no other research has previously analyzed micro blog data to identify needs of (potential) customers explicitly. The most comparable research is Misopoulos et al. (2014), who conducted a sentiment analysis with Twitter data to aggregate customer needs for the airline industry. Also based on Twitter, the researchers use sentiment analysis methods, whereas our work applies Machine Learning approaches to identify whether a tweet contains a need or not. As a result, we are able to train a model which can be later applied to unseen data sets—and therefore be reused without any additional effort.

---

[1] https://about.twitter.com/company, received on 11-09-2015

## 2.3      E-Mobility as Evaluation Domain

For testing our approach in an application domain, we require candidate domains to be both dependent on fast and ongoing monitoring of arising needs and rich in micro blog traffic. The domain of electric mobility (e-mobility) fulfills both our requirements. It is defined as "a highly connective industry which focuses on serving mobility needs under the aspect of sustainability with a vehicle using a portable energy source and an electric drive that can vary in the degree of electrification" (Scheurenbrand et al. 2015, p. 9). Recent studies (Pfahl et al., 2013; Sierzchula et al., 2014) highlight the relevancy of innovative services to foster the acceptance of e-mobility in society. However, creating ideas for, conceptualizing new offerings and providing a suitable service ecosystem in that space is still a challenge (Hinz et al., 2015; Stryja et al., 2015). Therefore, the systematic identification of customer needs that can be addressed via new (service) offerings is fundamental in this domain— and provides a relevant context for our study. At the same time, due to the currency and popularity of the topic, the number of micro blog data is enormous: based on relevant keywords determined by an expert workshop and a list of the top-selling electric vehicles in Germany we observed over 600.000 tweets within 6 months.

We further have to narrow down the domain to a geographical area with a coherent set of laws and regulations, markets as well as socio-economic conditions. In addition, we require micro blog data in a unique language, as we need consistent semantics to analyze. As a result of these requirements on the domain languages like English and Spanish—which can not be related to one region—are not suitable. Because of our familiarity with German, we focus on the German-speaking region. In addition to that, it fits into the agenda of the German government with its distinct goal to reach one million electric cars by 2020 (Bundesregierung, 2010) .

## 3   Method and Study

As highlighted previously, the study reported in the paper is part of the larger Needmining project. In order to structure our research in a rigorous way, we follow the DSR approach as presented by Kuechler and Vaishnavi (2008). In order to understand the problem related to the elicitation of customer needs out of mirco blogs, we conducted a series of expert interviews, an expert workshop as well as an intensive literature analysis. Having a clear understanding of the domain of e-mobility, the term "customer needs" and the various methods of micro blog data are the foundation for our work in this paper.

We develop an approach to classify micro blog data containing customer needs and test its feasibility—as a first DSR cycle. Table 1 presents the demarcation of the recent paper and the overall Needmining DSR project using the terminology proposed by March and Smith (1995).

| Scope | Objective | Criteria | Activities and Artifacts |
|---|---|---|---|
| Overall project (Needmining) | Improvement | Comparison to existing need elicitation methods | *Building* and *evaluating* the Needmining *artifact* and its instantiation as a comprehensive *software tool* |
| This work | Exaptation | Feasibility of automatic classification | *Building* an *approach* to identify wether or not a micro blog post contains a need with a resulting classification *instantiation* as a *classification tool* |

Table 1.         Overview of the general project and this work

In order to develop the approach for the classification of customer needs, we follow a five-step approach as presented in Figure 1. In a first step, we have to identify needs in micro blog data and thus, have to acquire micro blog data sets by using a publicly available source (*Data Retrieval*). Subsequently, it is important to gain an understanding for the micro blog messages at hand and their general structure (*Data Coding*)—which we achieve by applying a Descriptive Coding approach to exclude (for the purpose of this paper) irrelevant subsets of data (*Data Filtering*). Next, we obtain a categorization on whether or not a tweet contains a customer need by manually classifying an exemplary data set (*Data Labeling*). Finally, we build a supervised learning approach on this training set, applying different Machine Learning algorithms and the necessary preparations (*Preprocessing, Sampling, Classification*).

On a technical basis, we implement this method—as far as it concerns Data Retrieval, Data Coding, and Classification—as an artifact for acquiring, filtering, and analyzing micro blog data. The resulting artifact with its data pipeline is suited for processing both, historical and real time data.

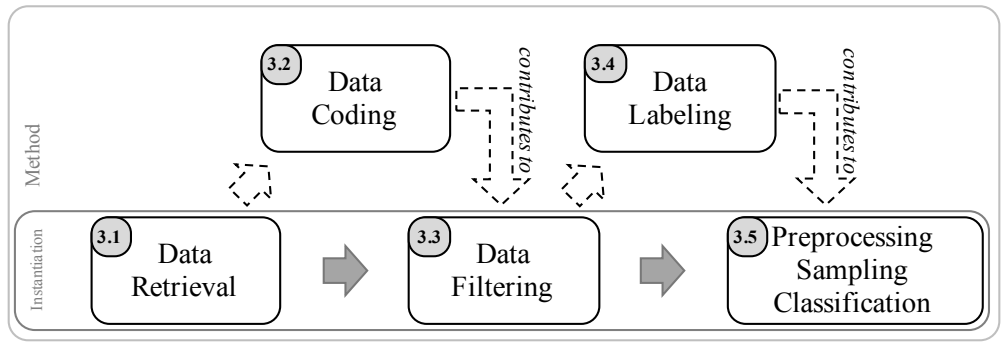

*Figure 1.     Overview of the Method Development*

## 3.1     Data Retrieval

First, we need to get access to large sets of micro blog data. As described previously, Twitter serves as a promising candidate source for data sets. However, Twitter does not offer research data sets of its micro blog data ("tweets") for public use. Instead, this research bases on data collected via Twitter's streaming API, which features an endpoint to fetch real-time data that matches pre-defined keywords, i.e. tweets that contain at least one of the pre-defined keywords. Using the streaming API enables us to receive *all* tweets matching the keywords (<400) as long as the fetched data does not exceed more that 1% of all tweets[2].

Our keyword set[3] consists of 33 words including (a) 17 German and 6 English generic terms[4] in the e-mobility domain, which are the outcome of a workshop with three professionals and (b) the names of the ten most popular electric vehicles based on the registration of new cars from 2011 until 2014 in Germany (Kraftfahrt-Bundesamt, 2014). Over a six-month-period from March to August 2015, we collect our data set via the Twitter streaming API. In addition to this data, we leverage the IBM Bluemix Insights for Twitter (IfT) service, which provides full access to a "statistically valid sample of at least 10% of all tweets" (so called "Decahose"[5]). We use the IfT service to add additional tweets from

---

[2] https://twittercommunity.com/t/best-solution-for-fetching-tweets-mentioning-hundreds-of-terms/28294, received on 11-09-2015

[3] Keywords (alphabetical, case-insensitive): bmw i3; e-tankstelle; eauto; ecar; egolf; electric mobility; electric vehicle; elektroauto; elektrofahrzeug; elektromobilitaet; elektromobilität; e-mobility; emobility; eup; fortwo electric drive; ladesaeule; ladesäule; miev; nissan leaf; opel ampera; peugeot ion; renault zoe; tesla model s.

[4] This research aims at analyzing German micro blog data for the time being. Because of the frequent use of certain English words (like e-Mobility or eCar) in otherwise German-speaking contexts, we include those words in the keyword set as well.

[5] https://gnip.com/sources/twitter/realtime/, received on 13-09-2015

*Kuehl et al. / Needmining*

the past (prior to the beginning of March) to our sample. In total, we acquire 645,226 tweets with tweets received from IfT making up 13.6% (absolute: 87,982) of all retrieved tweets.

## 3.2   Data Coding

To learn about content and composition of the obtained micro blog data, we manually code tweets using the descriptive coding approach as proposed by Saldaña (2012) and Krippendorff (2012). By coding a random set of the obtained tweets, we assign segments of data—in this case parts of a tweet's text—to a category, the code. This inductive approach helps to dynamically discover, merge, and delete codes—opposed to working with a fixed set of codes (deductive approach). Amongst others, we code sentiments, sentence structures, language components, named entities, and the type of tweet. The type of tweets refers to its general content and the author's intention for posting it, provided this is objectively reproducible. Additionally, we label the tweets according to the expressions of customer needs (*need-code*) identifying whether or not a customer posted a need. For the technical support of the coding, we use the application "maxQDA 11" as a software for qualitative data analysis.

We collaboratively perform the descriptive coding of 200 random German tweets by two researchers. The researchers assign the codes only if both researchers agreed upon it. There were no major disagreements in the coding process. Figure 2 shows an example for a coded tweet with the codes *institution/company*, *product*, *car*, *URL* and *feature/function*. After the explorative coding of 200 tweets, we identify 58 codes in total and reach a saturation in the identification of new codes. Then, we use the coding data to discover and test connections and relations between content and composition of a tweet and the occurrence of customer needs in it.

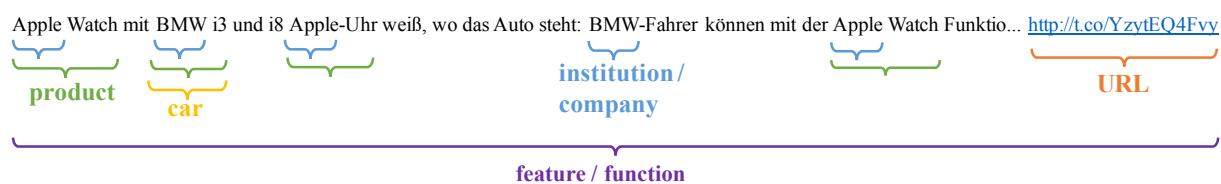

*Figure 2.        Example for a coded tweet in the Data Coding step*

## 3.3   Data Filtering

As our full data set covers more than 600,000 tweets, we need to reduce the data set to reach a higher share of tweets containing a need. First, we reduce our data set so that it only includes tweets that are expressed in German language (according to Twitter's metadata), which leaves us with 39,739 tweets (cf. Figure 3). Second, applying the insights of the coding, we analyze the correlation between identified customer needs and other codes: If we find a low confidence in the correlation of any code with the *need-code*, we might exclude data containing occurrences of this code for the *Data Labeling* and *Classification*—in order to reduce complexity. In fact, an association analysis yields one particularly noticeable correlation between the *need-code* and the code denoting the presence of a *URL*: Only in 3.64% of all tweets containing a URL (in total: 110), we could also find needs—supported by 11.5% observations in our sample. Therefore, we eliminate tweets with URLs from our total data set, reducing it by 91.5% from 39,739 to 3,368 tweets (cf. Figure 3). Third, we ensure that tweets with the same content are tagged and analyzed only once, and apply a deduplication method on our data set: The tweet's main content—its text without prefixes—is hashed with a hashing algorithm, and tweets with duplicate hashes are excluded, because they are either a retweet or occurred in both sources. Possible prefixes include "RT" to indicate a retweet and "@username" with username being the handle of a user the tweet is directed to. Figure 3 summarizes the results of our filtering. Starting from almost 40,000 tweets in German language, our coding analysis and deduplication filters reduces them to nearly 2,400 tweets which were subject to labelling in a next step.

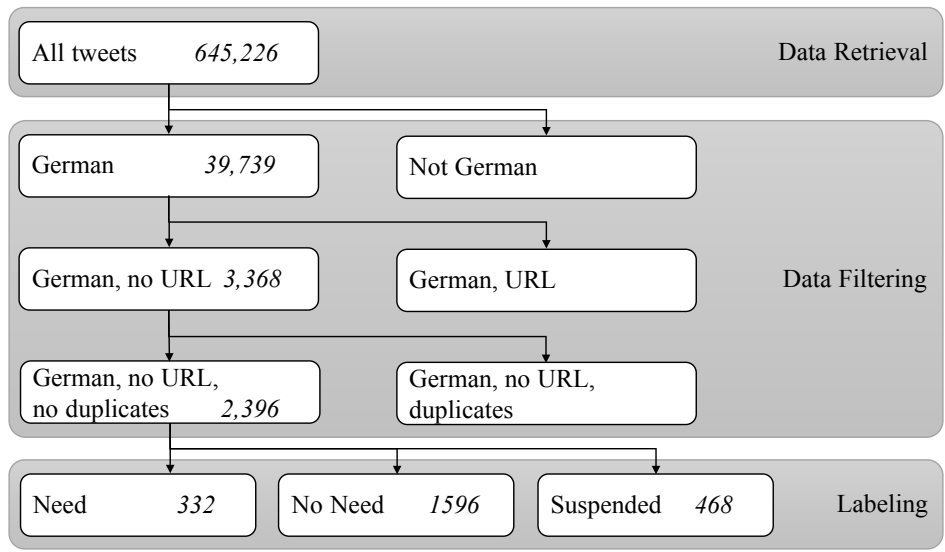

*Figure 3.     Number of used tweets in the individual steps*

## 3.4     Data Labeling

After filtering, the remaining 2,400 tweets need to be classified according to whether the message itself contains a customer need or not—in order to allow the building of a supervised learning model. To obtain this piece of information in an objective way, independent participants take part in a lab labeling session (Crowd Labeling). In this session, we instruct the participants to classify a set of tweets. They are incentivized and paid as described by Kvaløy et al. (2014), receiving a fixed payment. In total, 35 individuals[6] label all 2,396 relevant tweets in four 60-minute sessions with a maximum of 10 people each. Thereby, different participants classify each tweet three times. On average, it takes the participants about nine seconds per tweet to categorize it.

As a result (shown at the bottom of Figure 3), we define 332 tweets as *containing a need* (at least two of the taggers identify a need), 1,596 as *not containing a need* (none of the taggers identify a need), and the remaining 468 tweets as *suspend,* where only one of the taggers identify a need—as we consider this a disagreement regarding the existence of a need.

## 3.5     Preprocessing, Sampling and Classification

In order to identify the micro blog data containing needs, we have to take the following steps: We first preprocess the data to make it digestible for algorithms, explore different sampling procedures to deal with low need shares in the training set, and finally apply different standard Machine Learning algorithms.

*Preprocessing:* We use the text of a tweet as source for bag-of-words features for classification. We remove user names from the text, which gets then down-cased and tokenized. Subsequently, we stem[7] the tokens and remove stop words as well as too short tokens (shorter than two characters), before we convert the string tokens into a set of vector features by means of a transformation function representing Boolean word presence (see Figure 4).

---

[6] The participants were students in (Business) Engineering and received 10 € per hour.

[7] Stemming refers to the reduction of a word to its word stem. For stemming, we applied the Stanford Natural Language Processor.

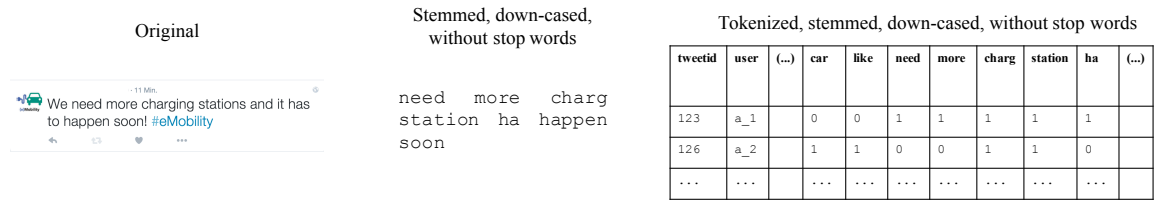

*Figure 4.    Simplified visualization of preprocessing steps for an English language tweet*

*Sampling:* Because of the imbalance of tweets containing needs (minority class with *332* instances) and tweets that do not contain needs (majority class with *1,596* instances), we consider to employ sampling steps to balance the data set for the model training. Balancing is necessary since classification algorithms only perform well with balanced data sets (Chawla, 2005). We separate the original data set into tweets containing needs (X) and tweets lacking needs (Y), as depicted in Figure 5. For sampling the training data, we use three different methods: *Undersampling* is randomly sampling instances from the majority class C (out of Y) to end up with enough instances to match the size of the data set A (out of X) (Rahman and Davis, 2013). *Oversampling* is getting additional instances from the minority class by duplicating random tweet instances to be able to work with more instances than originally available (Rahman and Davis, 2013). *Synthetic Minority Over-Sampling Technique (SMOTE)* creates new additional synthetic tweet instances (Ã) to match the training set of the majority class. In previous work, SMOTE-balanced data could be shown to achieve better classifier performance (Chawla et al., 2002). For each test set, we use the same distribution as the original data (B and E), without regarding already considered instances from the training sets.

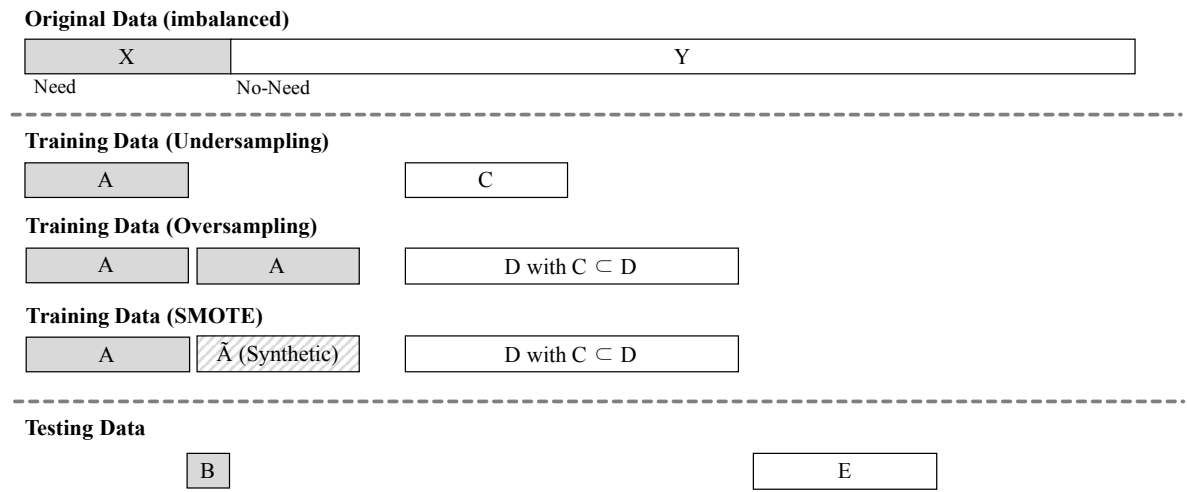

*Figure 5.    Visualization of different sampling techniques in original, training, and testing data*

*Classification algorithms:* In order to gain insights about their performance on our problem, we test several different popular classification algorithms (Michie et al., 1994) out of three groups; Bayes classifiers, Support Vector Machines, and tree-based classifiers (cf. Table 2).

| Group | Algorithm | Implementation |
|---|---|---|
| **Bayes classifiers** | Naïve Bayes | John and Langley (1995) |
| | Discriminative Multinomial Naive Bayes (DMNB) | Su et. al. (2008) |
| | Bayes Nets | Cooper and Herskovits (1990) |
| **Support Vector Machines** | Sequential Minimal Optimization based Support Vector Machines (SMO) | Platt (1998) |
| | SPEGASOS: Stochastic variant of Pegasos (Primal Estimated sub-GrAdient SOlver for SVM) | Shalev-Shwartz et al. (2007) |
| **Tree-based classifiers** | Random Trees | Breiman (2001) |
| | Random Forests | Breiman (2001) |

*Table 2.    Overview of applied algorithms and their implementation*

## 4  Results

This chapter presents the results from our study. All results are derived with a 10-fold cross-validation and a training-test-ratio of 2/3 to 1/3.

For quality measurement and comparison among different methods, we use the metrics of *accuracy* (the fraction of instances classified correctly), *precision* and *recall* (as defined by Van Rijsbergen (1979)) for each class as well as the *Area under the Receiver Operating Characteristic (ROC) curve (AUC)* (Fawcett, 2004). The ROC curve shows the performance of a (binary) classifier, plotting the true positive rate against the false positive rate for all different thresholds. The AUC is the area under that curve, reaching from 0.5 (random guess) to 1.0 (perfect classifier). Since it considers multiple factors, it is another well-suited indicator for the performance of a classifier (Powers, 2011).

In addition, we calculate the $F_\beta$-*score* as a weighted average for precision and recall (Van Rijsbergen, 1979):

$$F_\beta = (1 + \beta^2) * \frac{precision * recall}{(\beta^2 * precision) + recall}$$

For β=1, the score delivers a harmonic mean between precision and recall. However, β can be adjusted to reflect a higher emphasis of either precision (β<1) or recall (β>1)—depending on the intentions for using the classification model. This is relevant as the errors of miss-out on needs and fall-out (erroneously identifying needs) may entail different consequences for the innovation manager.

| Accuracy ▽ | Need Indicators | | | No Need Indicators | | Method | | Evaluation | | |
|---|---|---|---|---|---|---|---|---|---|---|
| | AUC | Precision (Need) | Recall (Need) | Precision (No Need) | Recall (No Need) | Sampling | Algorithm | $F_{0.5}$ | $F_1$ | $F_2$ |
| 85.270 | 0.621 | 0.685 | 0.268 | 0.865 | 0.974 | none | SVM | 0.522 | 0.385 | 0.305 |
| 84.647 | 0.627 | 0.614 | 0.292 | 0.867 | 0.962 | SMOTE | SVM | 0.503 | 0.396 | 0.326 |
| 84.647 | 0.586 | 0.696 | 0.193 | 0.854 | 0.982 | none | Bayes Net | 0.457 | 0.302 | 0.225 |
| 84.336 | 0.608 | 0.610 | 0.250 | 0.861 | 0.967 | none | SPEGASOS | 0.474 | 0.355 | 0.283 |
| 84.284 | 0.710 | 0.872 | 0.102 | 0.842 | 0.997 | none | DMNB | 0.348 | 0.183 | 0.124 |
| 84.232 | 0.762 | 0.671 | 0.166 | 0.850 | 0.983 | Oversampling | Random Forests | 0.417 | 0.266 | 0.195 |
| 84.180 | 0.669 | 0.575 | 0.310 | 0.869 | 0.952 | Oversampling | Bayes Net | 0.491 | 0.403 | 0.342 |
| 83.921 | 0.757 | 0.564 | 0.292 | 0.866 | 0.953 | Undersampling | Random Forests | 0.475 | 0.385 | 0.323 |
| 83.766 | 0.618 | 0.556 | 0.283 | 0.865 | 0.953 | SMOTE | SPEGASOS | 0.466 | 0.375 | 0.314 |
| 83.454 | 0.739 | 0.933 | 0.042 | 0.834 | 0.999 | none | Random Forests | 0.179 | 0.081 | 0.052 |
| 83.454 | 0.661 | 0.559 | 0.187 | 0.851 | 0.969 | SMOTE | Bayes Net | 0.399 | 0.280 | 0.215 |
| 83.402 | 0.743 | 0.833 | 0.045 | 0.834 | 0.998 | SMOTE | Random Forests | 0.186 | 0.086 | 0.056 |
| 83.402 | 0.708 | 0.929 | 0.039 | 0.833 | 0.999 | none | DMNB | 0.168 | 0.075 | 0.048 |
| 83.351 | 0.637 | 0.526 | 0.337 | 0.872 | 0.937 | Oversampling | SVM | 0.473 | 0.411 | 0.363 |
| 83.247 | 0.541 | 0.610 | 0.075 | 0.837 | 0.990 | Undersampling | Bayes Net | 0.252 | 0.134 | 0.091 |
| 82.832 | 0.632 | 0.502 | 0.331 | 0.870 | 0.932 | Oversampling | SPEGASOS | 0.455 | 0.399 | 0.356 |
| 82.417 | 0.699 | 0.481 | 0.274 | 0.861 | 0.939 | SMOTE | DMNB | 0.418 | 0.349 | 0.300 |
| 81.328 | 0.588 | 0.426 | 0.244 | 0.856 | 0.932 | none | Random Trees | 0.371 | 0.310 | 0.267 |
| 81.276 | 0.688 | 0.445 | 0.355 | 0.871 | 0.908 | SMOTE | Naïve Bayes | 0.424 | 0.395 | 0.370 |
| 81.120 | 0.573 | 0.407 | 0.211 | 0.851 | 0.936 | SMOTE | Random Trees | 0.343 | 0.278 | 0.233 |
| 80.809 | 0.672 | 0.445 | 0.464 | 0.887 | 0.880 | Undersampling | SVM | 0.449 | 0.454 | 0.460 |
| 80.187 | 0.717 | 0.406 | 0.325 | 0.865 | 0.901 | none | Naïve Bayes | 0.387 | 0.361 | 0.339 |
| 79.357 | 0.687 | 0.420 | 0.524 | 0.896 | 0.850 | Undersampling | SPEGASOS | 0.438 | 0.466 | 0.499 |
| 77.593 | 0.589 | 0.334 | 0.304 | 0.858 | 0.874 | Oversampling | Random Trees | 0.328 | 0.319 | 0.310 |
| 75.622 | 0.677 | 0.337 | 0.431 | 0.874 | 0.824 | Oversampling | DMNB | 0.353 | 0.378 | 0.408 |
| 74.015 | 0.630 | 0.322 | 0.461 | 0.877 | 0.798 | Undersampling | Random Trees | 0.343 | 0.379 | 0.424 |
| 61.826 | 0.715 | 0.273 | 0.729 | 0.913 | 0.595 | Oversampling | Naïve Bayes | 0.312 | 0.397 | 0.546 |
| 60.270 | 0.706 | 0.264 | 0.729 | 0.911 | 0.576 | Undersampling | Naïve Bayes | 0.302 | 0.387 | 0.539 |
| 17.324 | 0.562 | 0.172 | 1.000 | 1.000 | 0.001 | Undersampling | DMNB | 0.207 | 0.294 | 0.510 |

*Table 3.        Results of different classification algorithms*

Table 3 shows an overview of the results of different evaluations sorted by classification accuracy. For each analysis, the table states the quality of classification into the two classes of tweets containing needs ("need") or lacking needs ("no need")—derived for the test set based on a certain combination of training set sampling and classification algorithm. As an example, the first model in Table 3 (SVM algorithm without training data sampling) provides an overall accuracy of 85.3% on the classification, detecting 26.8% of all tweets with needs (recall need) and fall-out of 31.5% (1-0.685) for tweets that in fact do not contain needs.

The importance of focus on precision and recall in fact depends on the objective of innovation managers: In case they need to not miss out on any articulated need, they would go for higher recall—at the expense of having to evaluate more tweets classified as "need", but actually not containing one (false positives). In case they are interested in spending the least effort in further evaluating those tweets that get classified as needs, they would go with higher precision—at the expense of not capturing some potentially interesting micro blog information (false negatives). To reflect different inclinations for both types of error, we also show the $F_\beta$-scores for $\beta \in \{0.5; 1; 2\}$.

In a next step, we evaluated the different models and chose one—based on the different objectives the innovation manager may pursue (cf. Table 4):

- If we look from an *overall statistical point of view*, AUC is generally accepted as a meaningful classification indicator. AUC ranges between 0.5, which would be as good as a random guess, and 1.0, which would denote a perfect classifier. The best-performing method in this case is the combination of Oversampling and Random Forests, resulting in an AUC of 0.762, which is a fair result (cf. Metz 1978).
- If aiming for the *highest possible precision* (and thus lowest fall-out rate), the combination of no sampling and random forests delivers a very good result of 93.3%—at the price of only uncovering 4.2% of all need tweets.
- If we are aiming for the *highest possible recall*, the combination of Undersampling and Naïve Bayes achieves a recall result of 72.9%—but with a poor precision of 26.4%. A perfect recall of 100% is achieved by a combination of Undersampling and DMNB—but simply by assigning all tweets to the needs-class (and therefore not being practical at all).
- The best *balanced compromise between precision and recall* is a combination of Undersampling and SPEGASOS, which classifies more than half of the tweets wrongly (precision 42%) and uncovers half of all need tweets (recall 52%). Shifts of β would slightly increase the F-scores and recommend different algorithms.

| Objective | Evaluation Indicator | | Recommended Model | |
|---|---|---|---|---|
| | Key Indicator | Maximum Value | Actual Value | Sampling | Algorithm |
| Only regard tweets with high probability of containing a need—without concerns that only a fraction of all tweets with needs are identified | Precision (Need) | 1 | 0.933 | None | Random Forests |
| Uncover as many tweets with needs as possible—even if many of them don't actually contain needs | Recall (Need) | 1 | 0.729 | Undersampling | Naïve Bayes |
| Compromise between precision and recall | $F_1$-Score | 1 | 0.466 | Undersampling | SPEGASOS |
| Compromise between precision and recall—with focus on precision | $F_{0.5}$-Score | 1 | 0.522 | None | SVM |
| Compromise between precision and recall—with focus on recall | $F_2$-Score | 1 | 0.546 | Oversampling | Naïve Bayes |

*Table 4.    Model evaluation based on different application objectives*

There are options to further improve the models. For instance, we aim to increase the model performance by applying different preprocessing techniques like n-grams. Nonetheless, we can already select one of the models and apply it to unseen sets of Twitter data, which then allows for automatically identifying tweets containing needs with a high accuracy.

However, it should be noted, that our analysis is specific to the domain of e-mobility. An application to other domains without renewed training in that domain may result in a poorer classification. Nevertheless, we have shown how such models can be trained and applied to a specific domain to deliver valuable insights about micro blog data containing customer needs.

# 5    Conclusion and Outlook

In this paper, we analyze the feasibility to accurately and automatically classify micro blog data as to whether or not it contains customer needs. Even with standard Machine Learning algorithms, accuracies greater than 85% can be achieved. Certainly, there is a tradeoff to be made: with improved discovery of needs (need recall), also the number of erroneously identified needs tends to rise.

The proposed approach certainly has some limitations: First, we obtain the analyzed data set of micro blog data by using keyword-based streaming, i.e. possibly relevant data sets might be excluded based on a lack of appropriate keywords. Second, the modelling was done on a rather small data set (n=2,396) compared to the whole universe of tweets. Third, and most importantly, we tested the approach for one particular domain (e-mobility in Germany) only.

The managerial implications, however, are already valuable. Applying a fairly automated classification analysis to a set of publicly available data can lead innovation managers to find those "needles in the haystack" that contain valuable need information—and to do this on an ongoing basis. We can imagine a Needmining information system that presents innovation managers a selection of "need tweets" each morning. With the current state of the artifact, still manual work will be required to evaluate the expressed needs—but future work will help to make this task more effective by more granularly identifying and selecting needs. If this need information can be successfully translated into service innovations remains to be investigated as well.

On a theoretical level, we are able to present a continuous method to create a need-classification artifact for micro blog data by applying Machine Learning concepts to a problem in the field of service design.

A number of future research tasks are obvious: In parallel to further improving model quality, we need to apply the acquired classification model to huge data sets in order to identify possible needs of thousands of users. Then the value of the work can be tested in concrete use cases within the e-mobility industry. In addition, the method has to also be evaluated for different domains. And finally, in our current Needmining project, research is needed on whether we can retrieve more granular need information from micro blog data than a pure binary information whether the data does contain need information or not: An ample field for academically promising and industry-relevant future work lies ahead.